# Ceres as seen by VIR/ Dawn: spectral modeling and laboratory measurements suggest altered and pristine silicates within carbon chemistry

Andrea Raponi[1][0000-0003-4996-0099], Marco Ferrari[1][0000-0002-7447-6146], Maria Cristina De Sanctis[1][0000-0002-3463-4437], Eleonora Ammannito[2][0000-0001-6671-2690], Mauro Ciarniello[1][0000-0002-7498-5207], Filippo Giacomo Carrozzo[1][0000-0002-0391-6407]

[1] Institute for Space Astrophysics and Planetology-INAF, Via Fosso del Cavaliere, 100, 00133, Rome, Italy
[2] Italian Space Agency, ASI, Via del Politecnico snc, 00133, Rome, Italy
andrea.raponi@inaf.it

**Abstract.** The NASA/Dawn mission has acquired an unprecedented amount of data from the surface of the dwarf planet Ceres, providing a thorough characterization of its surface composition. The current favorite compositional model includes a mixture of ultra-carbonaceous material, phyllosilicates, carbonates, organics, Fe-oxides, and volatiles, as determined by the Dawn/VIR-IR imaging spectrometer, with the constraints on the elemental composition provided by the Dawn/GRAND instrument. The recent calibration of the VIS channel of the VIR spectrometer provided further constraints on the overall compositional model, suggesting another type of silicate not previously considered. This work reviews the VIR-VIS calibration process, and the resulting compositional model of Ceres surface considering both VIS and IR channels. Laboratory experiments are needed for a better understanding of the global (VIS-IR) shape of Ceres average surface, and in particular for the interpretation of the emergent absorption at 1μm. Current and future experiments are also discussed in this work.

## 1  Ceres' surface composition

Before the NASA/Dawn space mission, spectrophotometric ground observations of Ceres have revealed a complex mineralogy. The low albedo and the relatively flat visible/near-IR (0.4–2.5 μm) spectrum suggest the presence of a dark component like magnetite. Absorption bands at 3.3–3.4 μm and 3.8–4.0 μm have been interpreted by carbonates like dolomite, calcite, and magnesite. A narrow absorption at 3.06 μm has been attributed to different species, like ice frost, $NH_4$-bearing minerals, irradiated organics, crystalline water ice, cronstedtite, and brucite [1 and references therein], and thus its interpretation remained unclear until the NASA Dawn mission arrival. Dawn's payload includes the Visual and InfraRed mapping spectrometer (VIR), the Framing Camera (FC) with seven wavelength filters, and the Gamma Ray and Neutron Detector (GRaND). The three instruments were complementary in inferring the surface properties of Ceres. The VIR imaging spectrometer [2] is composed of two channels: the VIS channel covers the spectral range 0.25–1 μm, with a spectral sampling of ~2 nm, and the IR channel covers the spectral range 1–5 μm, with a spectral sampling of ~10 nm. The first hyperspectral images from the VIR-IR channel revealed a surface composed of a dark component (carbon and/or magnetite), and Mg-carbonates such as dolomite, confirming some of the previous studies based on ground observations. A newly discovered strong absorption at 2.7 μm was assigned to Mg-phyllosilicates such as antigorite, and a better characterization of the 3.06-μm absorption allowed us to assign it to $NH_4$-phyllosilicates [3]. Then, the GRaND instrument provided new constraints. Based on its elemental data, in particular the abundance of C, H, K, and Fe, [4,5] it was established that the dark material that makes up most of Ceres' surface composition should be rich in carbon, resembling carbonaceous chondrite material [6,7]. Recent improvement in the VIS channel calibration [8] allows for extending the range of the VIR instrument to shorter wavelengths (0.25–1.0 μm). Further calibration refinement was performed by Raponi et al. [1]. The resulting VIR data provided new constraints on Ceres' mineralogy.

## 2  VIR Calibration Refinement

Calibrated VIR online data (available through the Planetary Data System (PDS) archive at: https://sbn.psi.edu/pds/resource/dawn/dwncvirL1.html) are affected by the odd-even effect and systematic artifacts in the IR channel, and spurious spectral variations due to the detector temperature in the VIS channel. These can be corrected by methods implemented respectively by Carrozzo et al. [9] and Rousseau et al. [8]. The so calibrated data still present fictitious slopes which can be corrected by applying a multiplicative factor to both VIS and IR channels. Here we summarize the main steps requested to produce such correction vectors on the base of ground observations.

1. Ground observations of Ceres, which are mutually consistent, have been collected [1 and references therein], excluding out-of-statistics spectra.
2. For each ground full-disk observation (point 1), full-disk reflectance was converted to bidirectional reflectance in standard viewing geometry (incidence angle = 30°, emission angle = 0°, phase angle = 30°) by means of Hapke modeling [10] according to the photometric parameters derived by [11].
3. A smoothed average spectrum of ground observations has been calculated.
4. The average spectrum from VIR data photometrically corrected to standard viewing geometry [11] has been calculated, after artifact removal and photometric correction.
5. The ratio between the average spectrum from ground observations (point 3) and the average spectrum obtained from VIR data (point 4) has been calculated.
6. This ratio spectrum has been resampled for both VIS and IR channels. The two resulting vectors can be both used as multiplicative correction factors on every single VIR (VIS or IR) spectrum.

The so obtained correction factors (Fig. 1a) have been validated by performing a ratio between ground and VIR observation of Arcturus star spectra. That star has been chosen as calibration reference for its relatively brightness in both VIS and IR spectral ranges. However, given the high noise level, being a point like source, it cannot be used to produce independent ground correction vectors but just for a validation of that produced in the way discussed above.

## 3   Spectral Modeling

Merging the global average spectrum obtained in the VIS and IR channels, a single global average spectrum of Ceres in the range of 0.25–5.0 µm can be obtained (Fig. 1b). The thermal emission affecting the spectrum longward of 3.5 µm was removed as discussed in Raponi et al. [1] (Figure 1).
To obtain information on the abundance of the surface minerals, Raponi et al. [1] used a quantitative spectral analysis of the composition using Hapke's radiative transfer model [10]. Similar modeling has been described by De Sanctis et al. [3] and Marchi et al. [6]. The spectral range longward of 4.1 µm has been discarded because it is affected by a high noise level after the subtraction of the thermal emission. The average reflectance spectrum was modeled starting with minerals that have already been discussed in De Sanctis et al. [3] (see Table 1 in [3]). However, to account for the spectral shape in the VIS channel, Raponi et al. [1] added the spectrum of lizardite measured by Hiroi and Zolensky [12] who heated the sample in vacuum at 600°C to remove water contaminating the sample. Such a spectrum presents a peak close to 0.8 µm, similar to Ceres' average spectrum (Fig. 1b). However, it is significantly different from the lizardite spectrum measured at ambient conditions. Along with dehydration caused by the reduction in the water absorption at 3 µm, the high temperature triggered metamorphic reactions that altered its mineralogy and structure. The extent of this alteration deserves further investigation.
Marchi et al. [6] and Kurokawa et al. [7] found that a large fraction of Ceres surface should be composed of material similar to carbonaceous chondrites (CC), when taking into account both the IR global average spectrum and the elemental composition from the GRaND measurements. In particular, the best fits were obtained by adding the CC Ivuna (CI type) or LAP (CM type) to the modeled mixtures in [6] while the best fits were obtained using Ivuna (CI type), Tagish Lake (CM type), or Cold Bokkeveld (CM type) in [7]. However, analysis by Raponi et al. [1] revealed that the above-mentioned CC cannot match both the VIS and IR regions of the average spectrum of Ceres, and in particular the broad absorption at 1-µm (Fig. 1b). Instead, Raponi et al. [1] considered MAC 87300 (CM type), which was mentioned in a previous study [13].

## 4      Discussion and Conclusion

Raponi et al. [1] suggested the heated lizardite of Hiroi and Zolensky [12] as a possible analog. Lizardite, like antigorite, is a polymorph of serpentine; however, heating at 600 °C can lead to partial dihydroxylation and structural change, and thus to one or more different silicates. The fit with those thermally processed silicates does not necessarily imply that Ceres' surface experienced similar temperatures, which is not reasonable indeed. Instead, these silicates may be of primary origin, i.e., they were accreted from planetesimals that were thermally processed. Further laboratory measurements are needed to investigate the properties of these silicates, whose applicability to Ceres, if confirmed, would bring new constraints on Ceres' origin. Moreover, the need for a different carbonaceous chondrite (MAC 87300) than those considered in previous works, requires further analysis to explain which specific mineral phases determine its spectral shape and match with Ceres' spectra.

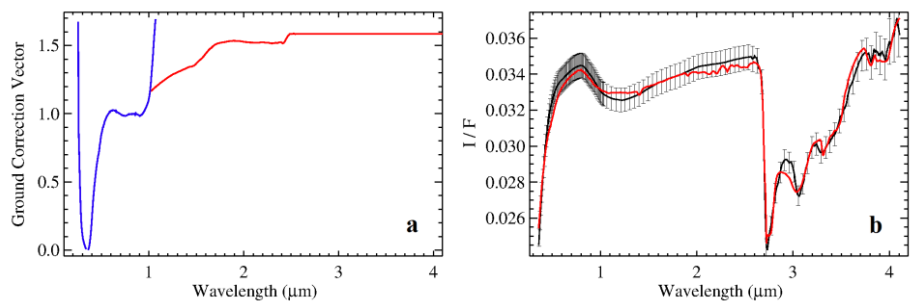

**Fig. 1.** Panel a: ground correction vectors for VIS (blue) and IR (red) channels to be used as a divider for each VIR spectrum. Panel b: average spectrum of Ceres surface after ground correction (black), and the best fit (red) as modeled by Raponi et al. [1].